\documentclass[12pt,nofootinbib,floatfix,onecolumn,preprintnumbers]{revtex4}

\pdfoutput=1
\usepackage{amsmath,amssymb}
\usepackage[utf8]{inputenc}
\usepackage{graphicx}
\usepackage{cancel,float}
\usepackage{soul}
\usepackage[usenames,dvipsnames]{color}
\usepackage[colorlinks=true]{hyperref}
\usepackage{subfig}
\usepackage{ulem}

\begin{document}
%%%%%%%%%%%%%%%%%%%%%%%%%%%%%%%%%%%%%%%%%%%%%%%%%
\title{Lepton dark matter portal in the inert Zee model}
\author{Alexandra Gaviria}%
\email{alexandra.gaviria@udea.edu.co}
\author{Robinson Longas}%
\email{robinson.longas@udea.edu.co}
\author{Andr\'es Rivera}%
\email{afelipe.rivera@udea.edu.co}
\affiliation{
Instituto de F\'isica, Universidad de Antioquia, Calle 70 No. 52-21, A.A. 1226, Medell\'in, Colombia}
\date{\today}

%%%%%%%%%%% ABSTRACT %%%%%%%%%%%%%%%%%%%%%%%%%%%%%%%%%%
\begin{abstract} 
The inert Zee model is an extension of the Zee model for neutrino masses. This new model explains the dark matter relic abundance, generates a one-loop neutrino masses and forbids tree-level Higgs-mediated flavor changing neutral currents. 
Although the dark matter phenomenology of the model is similar to that of the inert doublet model, the presence of new vector-like fermions opens the lepton portal as a new dark matter annihilation channel.
We study the impact of this new portal in the low-mass regime and show the parameter space allowed by direct and indirect searches of dark matter. Remarkably, the region for $m_{H^0} \lesssim$ 70 GeV is recovered for $\lambda_L \lesssim 10^{-3}$.
We also show that future experiments like LZ and DARWIN could probe a large region of the parameter space of the model.
\end{abstract}

%%%%%%%%%%% INTRODUCTION %%%%%%%%%%%%%%%%%%%%%%%%%%%%%
 \maketitle
\section{Introduction}
\label{sec:introduction}

Astrophysical observations suggest that the presence of dark matter (DM) in the universe is unquestionable. The latest data from the Planck experiment \cite{Aghanim:2018eyx} indicate that about $27\,\%$ of the Universe is composed of DM. However, its nature remains unknown since the Standard Model (SM) of particle physics does not fill the properties of a DM candidate.
Among different proposals beyond the SM to explain the DM problem, the Weakly Interacting Massive Particle (WIMP) is one the most popular since such a candidate connects the DM scenario and the SM through weak interactions. 
One of the simplest SM extensions that contain a WIMP is the Inert Doublet Model (IDM) \cite{Deshpande:1977rw}. 
In that framework, a second $\rm{SU(2)_L}$ scalar doublet is added to the SM and the stability of the lightest scalar state is ensured by the ad-hoc conservation of a $Z_2$ discrete symmetry. The model contains two viable DM mass regions that satisfy the relic abundance and direct detection (DD) limits \cite{Cirelli:2005uq,Barbieri:2006dq,Honorez_2007,Hambye_2009}, one above $m_{\text{DM}}\gtrsim 500$ GeV and the other one near the Higgs resonance $m_{\text{DM}}\sim m_h/2$. In the latter case, the dark sector communicates with SM through the Higgs portal. However, the same scalar coupling controls the DM annihilation into SM particles as well as the DM-nucleon scattering, constraining almost the entire region of the parameter space in the lack of DD spin-independent cross section limits.

On the other hand, further evidence of physics beyond the SM is given by the non-zero neutrino masses. If neutrino masses arise by radiative mechanisms \cite{Ma:1998dn,Bonnet:2012kz,Sierra:2014rxa}, it may be thought that these masses are related to the DM problem and, consequently, both could be originated at the TeV scale. In this direction, models with one-loop radiative neutrino masses and viable dark matter candidates have a complete classification given in Refs. \cite{Law:2013saa,Restrepo:2013aga,Yao:2017vtm,Carvajal:2018ohk}. In particular, the Inert Zee Model (IZM) is one of these realizations \cite{Longas:2015sxk} (cataloged as T1-ii-A model with $\alpha=-2$ in \cite{Restrepo:2013aga}). In the IZM, two vectorlike leptons, a singlet and a doublet of ${\rm SU(2)_L}$, and two scalar multiplets, a ${\rm SU(2)_L}$ doublet and a ${\rm SU(2)_L}$ singlet, are added to the SM. In addition, a discrete $Z_2$ symmetry, in which all the new fields are odd, is imposed to be unbroken providing a scalar DM candidate. 
The DM phenomenology of the IZM is quite similar to that of the IDM. In the high mass regime, when the particles not belonging to the IDM do not participate in the DM annihilation, $m_{\text{DM}}\gtrsim 500$ GeV. However, when these particles do take part of DM annihilation, the extra (not present in the IDM) coannihilation processes can modify this region allowing $m_{\text{DM}}\gtrsim 350$ GeV \cite{Longas:2015cnr,Klasen:2013jpa}\footnote{ Concerning the high mass region, it has been shown that the sensitivity of the Cherenkov Telescope Array (CTA) will be able to explore a large fraction of the allowed parameter space \cite{Queiroz:2015utg,Garcia-Cely:2015khw}.}. In contrast, the low mass region regime seems to be more affected because the DM candidate interacts with the SM through the new vectorlike leptons. This lepton portal can revive the low-mass regime of the IDM even if DD searches exclude the Higgs portal. Moreover, indirect detection (ID) experiments could be used to test the new portal.

In this work, we consider the IZM, and within this framework, we study the consequences of having new vector-like leptons in the DM phenomenology. In particular, we show that thanks to the lepton portal, a wide region of the IDM parameter space can be recovered in the low-mass regime. We show that for large Yukawa couplings, the relic abundance is correctly satisfied even when the Higgs portal is neglected. Furthermore, we show that lepton portal remains unconstrained from DD limits and we also present the restrictions given by ID experiments.

This paper is organized as follows: In section \ref{sec:IZM},
we present the model, including neutrino masses and DM. In section \ref{sec:Leptonportal}, we study the lepton portal and show the DM annihilation cross section. The numerical results are presented in section \ref{sec:Results}, and finally, we present our conclusions in section \ref{sec:conclusion}.

%%%%%%%%%%% THE MODEL %%%%%%%%%%%%%%%%%%%%%%%%%%%%%%%%%%
\section{The Inert Zee model}
\label{sec:IZM}

The IZM is an extension of the SM that includes
two vectorlike fermions, an ${\rm SU(2)_L}$-singlet $\epsilon$ and an ${\rm SU(2)_L}$-doublet $\Psi=(N, E)^{\text{T}}$. It also includes
two scalar multiplets, an ${\rm SU(2)_L}$-singlet $S^-$ and an ${\rm SU(2)_L}$-doublet $H_2=(H_2^+, H_2^0)^{\text{T}}$.
All of them are odd under a $Z_2$ symmetry, which in turn is used to avoid Higgs-mediated flavor-changing neutral 
currents at tree-level, forbid tree-level contributions to the neutrino masses 
and render the lightest $Z_2$-odd particle stable \cite{Longas:2015sxk}. 
The most general $Z_2$-invariant Lagrangian of the model contains the following new terms
\begin{align}
  \label{eq:Lmodel}
\mathcal{L}\supset -\left[ \eta_i\bar{L}_{i}H_2\epsilon + \rho_i \bar{\Psi}H_2 e_{Ri} + y \bar{\Psi} H_1 \epsilon 
+ f^*_i \overline{L^c_{i}} \Psi S^+  + {\rm h.c} \right] -\mathcal{V}\left(H_1,H_2,S\right)\,,
\end{align}
where  $L_i$ and $e_{Ri}$ ($i=1,2,3)$ are the SM leptons, doublets and singlets of ${\rm SU(2)_L}$ respectively, 
$\eta_i$, $\rho_i$ and $f_i$ are Yukawa couplings that control the new lepton interactions\footnote{We assume parity conservation in the new lepton sector. Thus, the term $\bar{\Psi}\gamma_5H_1\epsilon$ is neglected and the lepton mass matrix can be diagonalized by using only one real parameter.}, and the parameter $y$ is a coupling that leads to mixing among the $Z_2$-odd charged fermions. On the other hand, the scalar potential of the IZM is given by 
\begin{align}
\mathcal{V}\left(H_1,H_2,S\right) =  & \, \mu_1 H_1^{\dagger}H_1 + \frac{\lambda_1}{2}(H_1^{\dagger}H_1)^2 + \mu_2 H_2^{\dagger}H_2 + \frac{\lambda_2}{2}(H_2^{\dagger}H_2)^2+ \mu_S S^{\dagger}S + \lambda_S(S^{\dagger}S)^2 \nonumber \\  
& + \lambda_3 ( H_1^{\dagger}H_1 )( H_2^{\dagger}H_2 ) + \lambda_4 ( H_1^{\dagger}H_2 )( H_2^{\dagger}H_1 )  
 +\frac{\lambda_5}{2}\left[( H_1^{\dagger} H_2 )^2 + {\rm h.c.} \right]\nonumber \\
& + \lambda_6 (S^\dagger S) (H_1^{\dagger}H_1) - \lambda_7 (S^\dagger S)(H_2^{\dagger}H_2) -  \mu  \epsilon_{ab}\left[H_1^a H_2^b S
+ {\rm h.c.} \right]\,, 
\end{align}
where $\epsilon_{ab}$ is the ${\rm SU(2)_L}$ antisymmetric tensor, $\lambda_i$ and $\mu$ are scalar couplings which are assumed to be reals. After the electroweak symmetry breaking (ESB), the scalar Higgs doublet can be written as $H_1 = \left(0,(h + v)/\sqrt{2}\right)^T$, with $h$ being the Higgs boson and $v = 246$ GeV the vacuum expectation value (VEV). It is worth mentioning that $H_2$ does not develop a VEV in order to ensure the conservation of the $Z_2$ symmetry. 

The $Z_2$-odd scalar spectrum consists of a CP-even state $H^0$, a CP-odd state $A^0$ and two charged states $\kappa_{1,2}$ with masses 
\begin{align}
m^2_{H^0,A^0}&=\mu_{2}^{2}+\frac{1}{2}\left(\lambda_{3}+\lambda_{4}\pm\lambda_{5}\right)v^2\,,\\
m_{\kappa_1,\kappa_2}^2&=\frac{1}{2}\left\{m_{H^{\pm}}^2 + m_{S^\pm}^2 \mp [( m_{H^{\pm}}^2 - m_{S^\pm}^2 )^2 + 2\mu^2v^2]^{1/2}\right\}\,,
\end{align}
where $m_{H^{\pm}}^2 = \mu_2^2 + \frac{1}{2}\lambda_3 v^2$ and $m_{S^{\pm}}^2 = \mu_S^2 + \frac{1}{2}\lambda_6 v^2$.
The charged-scalar mixing angle $\delta$ is defined through $\sin{2\delta}=(\sqrt{2} \mu v)/(m_{\kappa_2}^2-m_{\kappa_1}^2)$. 
On the other hand, the $Z_2$-odd fermion spectrum involves two  charged fermions $\chi_{1,2}$ with masses given by
\begin{align}
m_{\chi_{1,2}} &=   \frac{1}{2}\left\{ m_{\Psi} + m_{\epsilon} \mp [( m_{\Psi} - m_{\epsilon} )^2 + 2y^2v^2]^{1/2}\right\},
\end{align}
and a mixing angle $\alpha$ which satisfies the relation $\sin{2\alpha} = (\sqrt{2} y v)/(m_{\chi_2} - m_{\chi_1})$. Also, there is a neutral Dirac fermion $N$, with a mass $m_N = m_{\Psi}$ such that $m_{\chi_1}\leq m_N\leq m_{\chi_2}$.

%%%%%%%%%%%%%%%%%%%%%%%%%%%%%%%%%%%%%%%%%%%%%%%%%%%%%%
\subsection{Neutrino masses}

\begin{figure}[t!]
  \begin{centering}
	\includegraphics[scale=0.7]{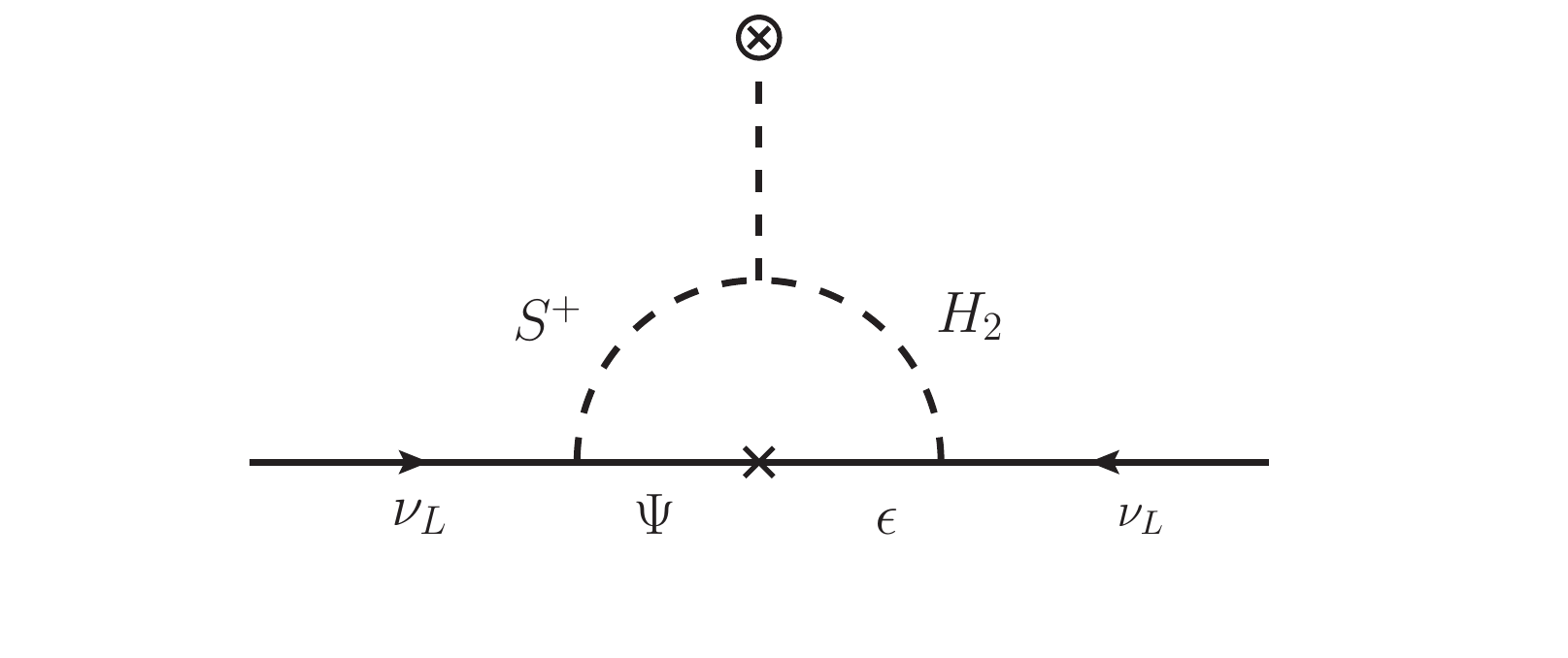}
	\caption{One-loop diagram leading to  neutrino Majorana masses. }
	\label{fig:neutrinomass}
	\end{centering}
\end{figure}

In the IZM, neutrino masses are generated at one-loop thanks to the scalar and fermion mixings and to the Yukawa interactions mediated by $\eta_i$ and $f_i$ (see Fig.~\ref{fig:neutrinomass}). The  neutrino mass matrix in the mass eigenstates is given by 
\begin{align}\label{eq:neutrinomass}
[M_{\nu}]_{ij} &= \zeta [\eta_i f_j + \eta_j f_i ],
\end{align}
where, 
\begin{align}
\zeta =&\frac{\sin2\alpha \sin2\delta}{64\pi^2}\sum_{n=1}^{2}c_nm_{\chi_n}I(m_{\kappa_1}^2,m_{\kappa_2}^2,m_{\chi_n}^2)\;,\,c_1=-1\;,\, c_2=+1, \\
&I(a,b,c)=b\ln(b/c)/(b-c) -  a\ln(a/c)/(a-c).
\end{align} 
Note that, because the flavor structure of $M_{\nu}$, the lightest neutrino is predicted to be massless in this model\footnote{Adding a fourth neutrino to the SM, it is possible to generated radiative Majorana masses for the three known neutrinos~\cite{Babu:1988ig}.}. 
The masslessness of the lightest neutrino entails several phenomenological consequences: $i)$ there is only single Majorana CP phase since the second phase can be absorbed by a redefinition of the massless neutrino field. $ii)$ The two remaining neutrino masses are determined by the solar and atmospheric mass scales: for normal hierarchy (NH) $m_1=0$, $m_2=\sqrt{\Delta m_{\text{sol}}^2}$ and $m_3=\sqrt{\Delta m_{\text{atm}}^2}$, while for inverted hierarchy (IH) $m_1=\sqrt{\Delta m_{\text{atm}}^2}$, $m_2=\sqrt{\Delta m_{\text{sol}}^2+m_1^2}\approx\sqrt{\Delta m_{\text{atm}}^2}$ and $m_3=0$. 
$iii)$ The amplitude for neutrinoless double beta decay \cite{Rodejohann:2011vc} presents a lower bound, which for the case of IH lies within the sensitivity of future facilities dedicated for that goal \cite{Reig:2018ztc}. 

In Ref.~\cite{Longas:2015sxk}, it was shown that, using Eq.~(\ref{eq:neutrinomass}) and the diagonalization condition\footnote{We work in the basis where the charged-lepton Yukawa matrix is diagonal.} $U^{\text{T}}M_{\nu}U ={\rm diag}(m_1,m_2,m_3)$ with  $U=VP$ and $P=\mbox{diag}(1,e^{i\phi/2},1)$~\cite{Agashe:2014kda}, it is possible to express five of the six Yukawa couplings $\eta_i$ and $f_i$ in terms of the neutrino low-energy observables. 
Consequently, the most general Yukawa couplings that are compatible with the neutrino oscillation data are given by
\begin{align}\label{eq:yuks-f-eta}
&\eta_i=|\eta_1|\frac{A_i}{\beta_{11}}\,, \hspace{1cm}f_i=\frac{1}{2\zeta}\frac{\beta_{ii}}{\eta_i}\,,
\end{align}
where 
\begin{align}
\beta_{ij} &= e^{i\phi} m_2V_{i2}^*V_{j2}^*+ m_3V_{i3}^*V_{j3}^*\,,\nonumber\\
A_{j} &= \pm\sqrt{-e^{i\phi} m_2m_3(V_{12}^*V_{j3}^*-V_{13}^*V_{j2}^*)^2e^{i2\text{Arg}(\eta_1)}}+\beta_{1j}e^{i\text{Arg}(\eta_1)}\,,\,\,\,  \mbox{for NH}\,, \label{eq:betaij1}\\
\beta_{ij} &= m_1V_{i1}^*V_{j1}^*+ e^{i\phi} m_2V_{i2}^*V_{j2}^*\,,\nonumber\\ A_{j} &= \pm\sqrt{-e^{i\phi} m_1m_2(V_{11}^*V_{j2}^*-V_{12}^*V_{j1}^*)^2e^{i2\text{Arg}(\eta_1)}}+\beta_{1j}e^{i\text{Arg}(\eta_1)}\,,\,\,\, \mbox{for IH}\,.
\label{eq:betaij2} 
\end{align}
This way, it is always possible to correctly reproduce the neutrino oscillation parameters in the present model. Finally, it should be noted from Eq. \eqref{eq:yuks-f-eta} that only the Yukawa coupling $\eta_1$ remains as a free parameter in the neutrino sector.

%%%%%%%%%%%%%%%%%%%%%%%%%%%%%%%%%%%%%%%%%%%%%%%%%%%%%%
\subsection{Dark Matter}
The conservation of the $Z_2$ symmetry ensures the stability of the lightest odd particle. In the IZM, as in the IDM, the pseudoscalar $A^0$ or the scalar $H^0$ can be the lightest state\footnote{Note that the neutral fermion $N$ can not play the role of the DM candidate since $m_{\chi_1}\leq m_N$.}. Without loss of generality we assume $H^0$ to be the DM candidate\footnote{The choice of $A_0$ as the DM candidate does not change significantly the phenomenology of the model \cite{Arhrib:2013ela} and regarding the lepton portal, it remains unchanged. The only effect is a phase factor ($\pm i$) in the Yukawa Lagrangian (eq.~\eqref{eq:Lmodel}) that not change the analysis.}.
Hence, the DM phenomenology of the IZM is expected to be similar to the one in the IDM in scenarios where the particles not belonging to the IDM ($\kappa_{1,2}$, $\chi_{1,2}$ and $N$) do not participate in the DM annihilation processes \cite{Longas:2015sxk,Longas:2015cnr}.  
Accordingly, the viable DM mass range for this scenario is divided into two regimes \cite{Barbieri:2006dq,Honorez_2007,Honorez:2010re,LopezHonorez:2010tb,Goudelis:2013uca,Garcia-Cely:2013zga,Arhrib:2013ela}: 

{\bf Low mass regime} ($53 \lesssim $  $m_{H^0}/{\rm GeV} \lesssim 75 $ ): 
In this mass range the main annihilation modes are through the Higgs $s-$channel exchange into light fermions (mainly $bb$ quarks) controlled by the quartic coupling $\lambda_L\equiv \frac{1}{2} (\lambda_{3} + \lambda_{4} + \lambda_{5})$ . Furthermore, LEP measurements give rise to the following constraints \cite{Lundstrom:2008ai}: $m_{H^0} + m_{A^0} > m_Z$, max$( m_{H^0} , m_{A^0} ) > $ 100 GeV and $m_{\kappa_{1,2}} \gtrsim70$ GeV.
It is worth mentioning that the combination of DD searches and the invisible Higgs decays exclude the region below $m_{H^0}\lesssim 55$ GeV; but we point out here that, because of the presence of the new lepton portal that will be described in next section, the IZM  allows to reproduce the correct relic abundance for $\lambda_L \sim 0$ and DM masses $m_{H^0}\lesssim 55$ GeV if the Yukawa couplings that mediate such annihilation, $\rho_i$, are of order of one (see section \ref{sec:Results}).
In that a case, the IZM would recover this region that is already excluded in the IDM.

{\bf High mass region} ($m_{H^0} \gtrsim 500$ GeV) : In this regime, the relic abundance depends strongly on the mass splittings between $H^0$, $A^0$ and $\kappa_1$. Indeed, a small splitting of at most $15$ GeV is required to reproduce the correct relic density implying that coannihilations between those particles must be taken into account.

In the intermediate mass region, $100 \lesssim m_{H^0}/{\rm GeV} \lesssim 500$, the gauge interactions become large so that it is not possible to reach the observed relic density, i.e. $\Omega_{H^0} < \Omega_{DM}$. Hence, this mass region has
been entirely excluded in the light of recent DD limits and relic density constraints. 

%%%%%%%%%%%%%%%%%%%%%%%%%%%%%%%%%%%%%%%%%%%%%%%%%%%%%%
\section{Lepton portal}
\label{sec:Leptonportal}
\begin{figure}[t!]
	\begin{centering}
		\includegraphics[width=0.5\textwidth]{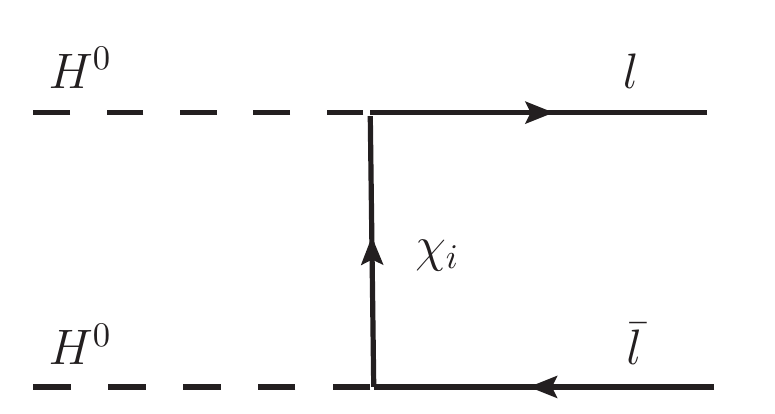}
		\caption{Feynman diagrams for annihilation of dark matter to SM particles, through $ t $-channel mediated by
			$ \chi_i $ with final states leptons $ \ell \bar {\ell} $. }
		\label{fig:channels}
	\end{centering} 
\end{figure}
The presence of charged vectorlike leptons in the IZM opens the lepton portal DM and, additionally to Higgs portal, becomes one of the main annihilation channels for the scalar DM in the low mass regime. The contributions for new $H^0$ annihilation into a SM lepton pair are mediated by the fermions $\chi_i$ as shown in Fig.~\ref{fig:channels} and the corresponding cross section depends on the Yukawa couplings $\eta_i$ and $\rho_i$ (see Eq. \eqref{eq:Lmodel}).
These couplings also generate charged lepton flavor violation processes (CLFV) and give contributions to electron-electric dipole moment (eEDM) which plays an important role in the DM relic abundance. As shown in a recent work \cite{Gaviria:2018cwb}, this kind of constraints are satisfied by taking $|\eta_1|\lesssim 10^{-2}$ in the neutrino NH and $|\eta_1|\lesssim 10^{-1}$ in the IH.
Also, taking $\sqrt{|\rho_1||\rho_2|}\lesssim 10^{-2}$, which is independent of the neutrino hierarchy since the Yukawa couplings $\rho_i$ do not take part of neutrino physics.
Finally, $\rho_3$ remains unconstrained by the CLFV limits.

For low values of the Higgs portal coupling, $\lambda_L$, it is possible to solve the Boltzman equations and compute the relic abundance of DM~\cite{Kolb:1990vq, Griest:1990kh},
\begin{equation}
\label{eq:elportal}
\Omega h^2 = \dfrac{1.07 \times 10^{9}x_f}{M_{p}\sqrt{g^*}\left(a+ \dfrac{3\, b}{x_f}+\dfrac{20\, c}{x_f^2} \right)}\,,
\end{equation}
where $x_f=m_{H^0}/T$, $g^{*}$ are the relativistic degrees of freedom at the freeze-out temperature and $M_p\approx 1.22 \times 10^{19}$ GeV is the Planck mass. The parameters $a, b$ and $c$ are obtained after computing the velocity annihilation cross-section which is given by\footnote{We have neglected the contributions involving the $\eta_i$ Yukawa because they are suppressed by CLFV processes \cite{Gaviria:2018cwb}.}
\begin{equation}
\label{eq:sv}
\langle \sigma v\rangle = a + b\, v^2 + c\, v^4 + \mathcal{O}(v^6)\,.
\end{equation}
The parameters $a,b,c$ are the s-wave, the p-wave and the d-wave coefficients respectively given by, 
\begin{eqnarray}
\label{eq:abc}
a &=& \dfrac{\rho_i^4}{4\pi m_{H^0}^2}\dfrac{m_{l_i}^2}{m_{H^0}^2}\dfrac{1}{(1+\mu)^2}\,,\nonumber\\
b &=& -\dfrac{\rho_i^4}{6\pi m_{H^0}^2}\dfrac{m_{l_i}^2}{m_{H^0}^2}\dfrac{1+2\mu}{(1+\mu)^4}\,,\nonumber\\
c &=& \dfrac{\rho_i^4}{60\pi m_{H^0}^2}\dfrac{1}{(1+\mu)^4}\,,
\end{eqnarray}
with $\mu=m_{\chi_i}/m_{H^0}$. Note that the s-wave and the p-wave contributions are helicity suppressed, whereas d-wave is the leading term for $m_{l_i}\rightarrow 0$ and becomes the dominant contribution in the early universe.  
It is worth mentioning that we have not considered the internal Bremmsstrahlung processes in the DM annihilation because its contribution is roughly one order of magnitude lower than the 2-body decays showed in Fig. \ref{fig:channels} \cite{Toma:2013bka, Giacchino:2013bta}.

%%%%%%%%%% Results %%%%%%%%%%%%%%%%%%%%%%%%%%%%%%%%%%
\section{Numerical results}
\label{sec:Results}
\begin{figure}[t!]
	\begin{centering}
		\includegraphics[scale=0.65]{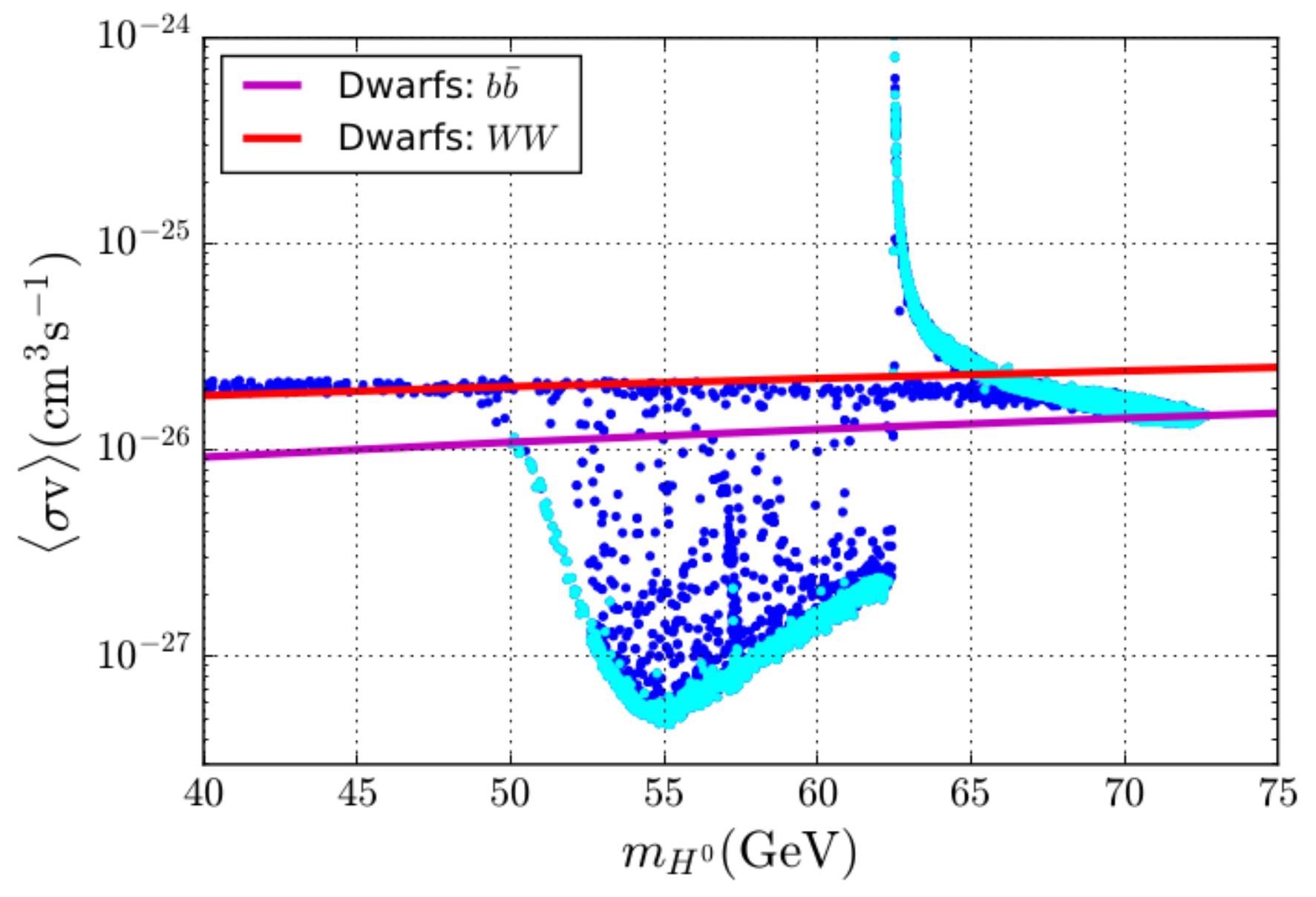}
		\caption{The present velocity averaged annihilation cross section as a function of the DM mass, $m_{H^0}$. The cyan points correspond to the IDM allowed parameter space, whereas the blue points has been recovered using the lepton portal present in the IZM. Also, it is shown the experimental limits for DM annihilation into $b\bar{b}$ and  $W^+W^-$ in dwarf spheroidal galaxies (dSphs)~\cite{Ackermann:2015zua}.}
		\label{fig:taus1}
	\end{centering} 
\end{figure}

In order to study the DM phenomenology of the IZM in the low-mass scenario, we have performed a random scan over the parameter space, varying the free parameters as 
\begin{align}
\label{eq:scanLFVlow2}
& 10^{-4} \leq | \eta_1 |,|\rho_1|,|\rho_2|,|\rho_3| \leq 3\;;\;\nonumber\\
& 100 \, {\rm GeV} \leq m_{A^0},\, m_{\kappa_1},m_{\chi_1}  \leq 500\, {\rm GeV}\;;\;\nonumber\\
&m_{\kappa_2} =  [m_{\kappa_1},500\,{\rm GeV}]\,;\;\\
& m_{\chi_2} =  [m_{\chi_1},500\, {\rm GeV}]\;;\;\nonumber\\
&40\, {\rm GeV} \leq m_{H^0}\leq 75\, {\rm GeV}\,.\nonumber
\end{align}
\begin{figure}[t!]
	\begin{centering}
		\includegraphics[width=0.65\textwidth]{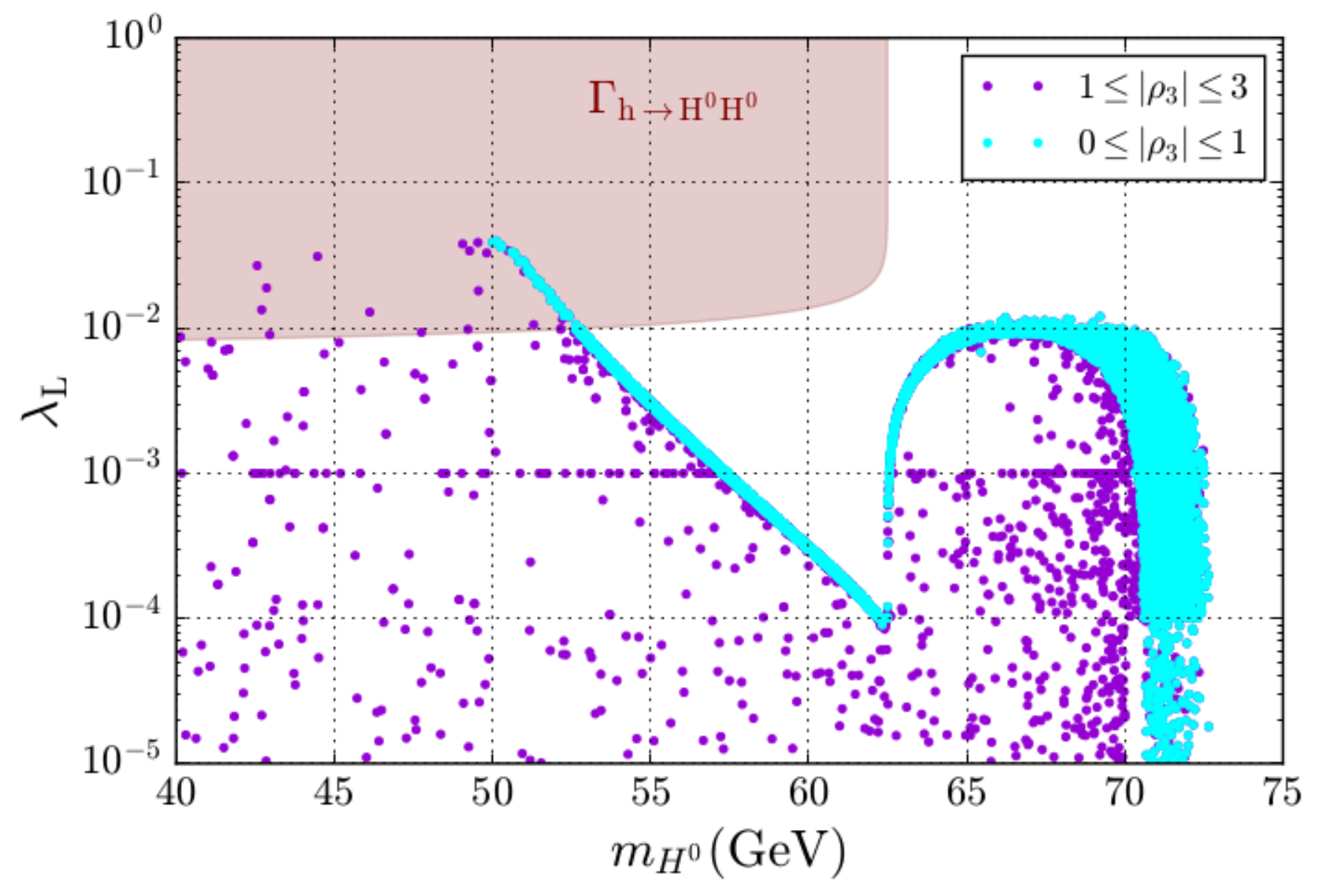}
		\caption{The available parameters space on the plane ($m_{H^0}$, $\lambda_L$). The purple and cyan regions are   compatible with the experimental value for the relic density.  Also, it is shown the constraints on the invisible branching of the Higgs boson (brown region) \cite{Belanger:2013xza}.}
		\label{fig:rho3}
	\end{centering} 
\end{figure}
We have implemented the model in~\texttt{SARAH}~\cite{Staub:2008uz,Staub:2009bi,Staub:2010jh,Staub:2012pb,Staub:2013tta}, coupled to the \texttt{SPheno}~\cite{Porod:2003um,Porod:2011nf} routines and, in order to obtain the DM relic density, we have used~\texttt{MicrOMEGAs}~\cite{Belanger:2006is}. 
We have checked the numerical results with the expression found in Eq. \eqref{eq:elportal} for the lepton portal limit and we have taken the points that fulfill the current value of the relic density $\Omega h^2 = (0.120 \pm 0.001)\; \text{to}\; 3\sigma$~\cite{Aghanim:2018eyx}. Besides, we ensure that the $S$, $T$ and $U$ parameters remain within the $3\sigma$ level \cite{Baak:2014ora}. Also, we have used the \texttt{FlavorKit}\cite{Porod:2014xia} of
\texttt{SARAH} to select those points that satisfy the CLFV constraints.
\begin{figure}[t!]
	\begin{centering}
		\includegraphics[scale=0.7]{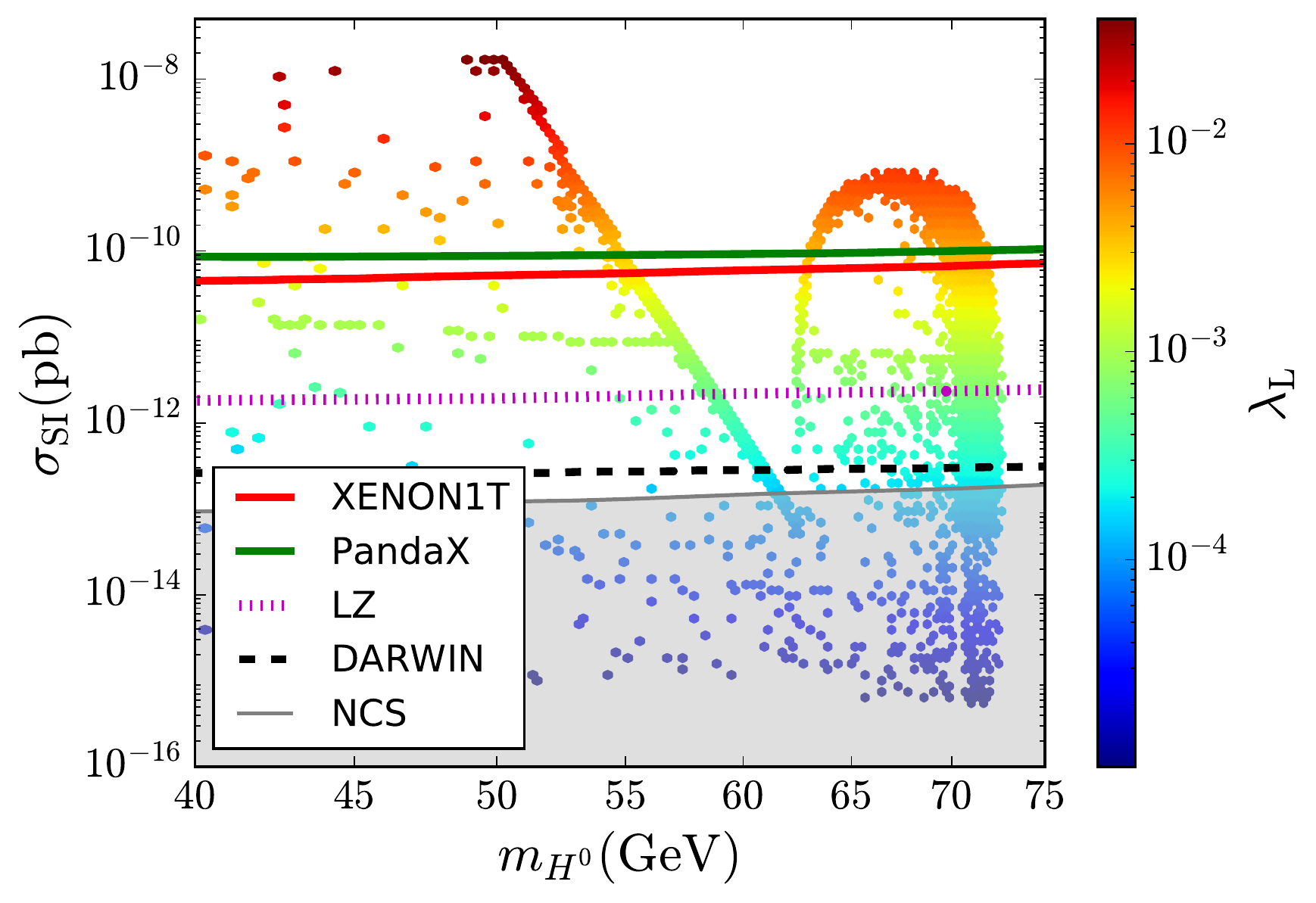}
		\caption{Spin-independent cross-section as a function of the DM mass $m_{H^0}$. The color bar shows the values for the Higgs portal coupling, $\lambda_L$. Also, it is is shown the current experimental constraints from XENON1T~\cite{Aprile:2018dbl}, PandaX~\cite{Cui:2017nnn} and prospects as LZ~\cite{Akerib:2018lyp}, DARWIN~\cite{Aalbers:2016jon}. 
			We also show the Neutrino Coherent Scattering (NCS)~\cite{Cushman:2013zza, Billard:2013qya}.}
		\label{fig:taus}
	\end{centering} 
\end{figure}
The results are showed in Figs.~\ref{fig:rho3}, \ref{fig:taus} and \ref{fig:taus1} respectively. Fig.~\ref{fig:rho3} shows the results on the plane ($m_{H^0}$, $\lambda_L$). All dots satisfy the relic abundance, the CLFV constraints and the oblique parameters. As can be seen, we find that for  $0 \lesssim |\rho_3| \lesssim 1 $ the parameter space of IDM (cyan region) is recovered and the DM is annihilated mainly to $b \bar{b}$ quarks through the Higgs portal. While, for $1 \lesssim |\rho_3| \lesssim 3 $ we obtain a new allowed window for the DM mass (purple region). In this region, DM is annihilated through the lepton portal  mediated by $\chi_i^\pm$  to  $b \bar{b}$ and $\tau \tau $. Note, for instance, that thanks to the lepton portal, the DM mass region for $m_{H^0}\lesssim 55$ GeV is allowed, even for small a scalar coupling ($\lambda_L \lesssim 10^{-3}$).
Also, the brown region shown the restriction on the Higgs invisible width decay taking the upper bound for $ \mathcal{B}_{\text{inv}}= 0.191 $ \cite{Belanger:2013xza},  this  constraint excludes values for $\lambda_L$ above $10^{-2}$, as is usual in the IDM. 

Fig \ref{fig:taus} shows the spin-independent cross-section as a function of the DM mass, $m_{H^0}$. Red and green horizontal lines represents the experimental limits coming from  XENON1T \cite{Aprile:2018dbl} and PandaX searches \cite{Cui:2017nnn}. Also, the prospects for future searches from LZ \cite{Akerib:2018lyp} and Darwin \cite{Aalbers:2016jon} are shown in magenta and black dashed lines. Note that the region for $m_{H^0}\lesssim 55$ GeV for $\lambda_L \lesssim 10^{-3}$  is allowed by the current and future DD searches, due to the presence of the lepton portal DM through the $\rho_3$ Yukawa coupling. However, below $\sigma_{SI}\lesssim 10^{-13}$ pb, the parameter space of the IZM would not be distinguishable from the neutrino floor \cite{Cushman:2013zza, Billard:2013qya} and it needs special analysis that is beyond the scope of this work.

Finally, in Fig \ref{fig:taus1} we show the velocity annihilation cross-section as function of the DM mass, $m_{H^0}$. The restrictions coming from FermiLAT \cite{Ackermann:2015zua} are represented by the continuous lines. As we can see, these restrictions exclude the mass region of the model for $m_{H^0} \lesssim 50$ GeV. However, between 50 GeV $\lesssim m_{H^0} \lesssim$ 70 GeV there is a viable window which is not included in the IDM (blue points). For this region, $m_{H^0}\lesssim$ 60 GeV, we have that the dominant channel is the lepton portal to $\tau \tau$ because the DM ($H^0$) does not has enough energy to produce a Higgs, while for $m_{H^0}\gtrsim$ 60 GeV the main channel is the annihilation to a $b  \bar{b}$ pair trough the SM Higgs.

%%%%%%%%%%%%%%%%%%%%%%%%%%%%%%%%%%%%%%%%%%%%%%%%%%%%%%
\section{Conclusions}
\label{sec:conclusion}
In this work, we have explored the DM phenomenology of the Inert Zee model in the low mass regime being compatible with neutrino and CLFV observables. We showed that in such a region, there exist two important portals for DM annihilation into SM particles: the Higgs portal, which depends of the scalar coupling $\lambda_{L}$ and the lepton portal, mediated by the Yukawa coupling $\rho_3$. When the Higgs portal dominates the annihilations, $ |\rho_3| \lesssim 1 $, the viable allowed DM mass region is around the Higgs mass, $m_{H^0}\sim m_h/2$ as in the case of the IDM. This takes place because the scalar coupling $\lambda_{L}$ enters in the annihilation channels and, at the same time, controls the DD spin-independent cross section as well as the Higgs invisible decays, which are well constrained. On the other hand, when the lepton portal dominates the DM annihilation, $1 \lesssim |\rho_3| \lesssim 3 $, the DD restrictions are evaded and a wide region of the DM mass, 50 GeV $\lesssim m_{H^0} \lesssim$ 70 GeV,  is still allowed. Moreover, in this region, the model is safe from CLFV processes and also satisfy the ID restrictions.

%%%%%%%%%%%%%%%%%%%%%%%%%%%%%%%%%%%%%%%%%%%%%%%%%%%%%%
\section*{Acknowledgments} 
We are grateful to Walter Tangarife for reading the manuscript and Oscar Zapata for enlightening discussions.
This work has been partially supported by the Sostenibilidad program of Universidad de Antioquia UdeA, CODI-E084160104 grant and by COLCIENCIAS through the grants 111565842691 and 111577657253. A.R was also supported by COLCIENCIAS through the ESTANCIAS POSTDOCTORALES program 2017.

\newpage
%%%%%%%%%%%%%%%%%%%%%%%%%%%%%%%%%%%%%%%%%%%
\bibliographystyle{h-physrev4}
\bibliography{references}
%%%%%%%%%%%%%%%%%%%%%%%%%%%%%%%%%%%%%%%%%%%%%%%%%
\end{document}